\documentstyle[prl,eqsecnum,aps]{revtex}

\begin{document}
\author{Jian-Qi Shen $^{1,}$$^{2}$ \footnote{E-mail address: jqshen@coer.zju.edu.cn}}
\address{$^{1}$  Centre for Optical
and Electromagnetic Research, State Key Laboratory of Modern
Optical Instrumentation, \\Zhejiang University,
Hangzhou SpringJade 310027, P. R. China\\
$^{2}$Zhejiang Institute of Modern Physics and Department of
Physics, Zhejiang University, Hangzhou 310027, P. R. China}
\date{\today }
\title{On the self-induced charge currents in electromagnetic materials
\\and its effects in the torsion balance experiment\footnote{This paper is
concerned with the self-induced charge currents and some related physically interesting topics.
It will be submitted nowhere else for the publication, just uploaded at the e-print archives.}}
\maketitle

\begin{abstract}
We concern ourselves with the {\it self-induced charge currents}
in electromagnetic materials and some related topics on its
effects in the present paper. The contribution of self-induced
charge currents of metamaterial media to {\it photon effective
rest mass} is briefly discussed. We concentrate primarily on the
torque, which is caused by the interaction of self-induced charge
currents in dilute plasma with interstellar magnetic fields,
acting on the torsion balance in the torsion balance experiment.
It is shown by our evaluation that the muons and alpha-particles
in secondary cosmic rays will contribute an effective rest mass
about $10^{-54}$ Kg to the photon, which is compared to the newly
obtained upper limit on photon rest mass in Luo's {\it rotating
torsion balance} experiment.
\end{abstract}
\section{Photon effective rest mass due to self-induced charge current}
In is well known that in some simple electromagnetic media such as
electron plasma, superconducting media and Lorentz dispersive
materials (inside which the time-harmonic electromagnetic wave is
propagating), the self-induced charge current density ${\bf
J}$\cite{Ho} is proportional to the magnetic vector potential
${\bf A}$. Thus the interaction term $\mu _{0}{\bf J}\cdot {\bf
A}$ in the Lagrangian density is therefore transformed into an
effective mass term $-\frac{1}{2}\frac{m_{\rm eff}^{2}c^{2}}{\hbar
^{2}}{\bf A}^{2}$ of electromagnetic fields\cite{Shen}, which is
analogous to the rest mass term in the Londons' electromagnetics
for superconductivity, Ginzberg-Landau superconductivity theory
and Higgs mechanism. Note that in these media, the magnetic
permeability\footnote{Note that here for the magnetic properties
in superconductors, we adopt the current viewpoint rather than the
magnetic-charge viewpoint (where the permeability can be viewed as
$\mu=0$).} $\mu=1$, so it is easy for us to obtain ${\bf J}$ (and
hence the {\it effective rest mass} of photon) by solving the
equation of motion of charged particles acted upon by
electromagnetic waves. However, for some artificial composite
electromagnetic metamaterials, which are described by, {\it e.g.},
the {\it two time derivative Lorentz material} (2TDLM) model (in
which both the electric permittivity and the magnetic permeability
are of frequency dependence and therefore of complicated form) and
some uniaxially (or biaxially) anisotropic media with permittivity
and permeability being tensors, the above formulation may be not
helpful in obtaining the {\it effective rest mass} of
electromagnetic fields. In these cases, however, we have shown
that the following formula\cite{Shen}
\begin{equation}
\frac{m_{\rm eff}^{2}c^{4}}{\hbar ^{2}}=(1-n^{2})\omega ^{2},
                       \label{eq1}
\end{equation}
which arises from the Einstein-de Broglie relation, can be
applicable to the above problem. In the expression (\ref{eq1}),
$n$, $\hbar$ and $c$ denote the optical refractive index, Planck's
constant and speed of light in a vacuum, respectively.

The {\it two time derivative Lorentz material} (2TDLM) model first
suggested by Ziolkowski and
Auzanneau\cite{Ziolkowski3,Ziolkowski4} is a generalization of the
standard Lorentz material model. Ziolkowski has shown that this
type of 2TDLM medium can be designed so that it allows
communication signals to propagate in the medium at speeds
exceeding the speed of light in a vacuum without violating
causality\cite{Ziolkowski}. For these artificial metamaterials,
the model encompasses the permittivity and permeability material
responses experimentally obtained\cite{Smith}. The \^{x}-directed
polarization and \^{y}-directed magnetization fields in such
materials would have the following forms\cite{Ziolkowski}
\begin{eqnarray}
\frac{\partial^{2}}{\partial t^{2}}{\mathcal
P}_{x}+\Gamma\frac{\partial}{\partial t}{\mathcal
P}_{x}+\omega^{2}_{0}{\mathcal
P}_{x}=\epsilon_{0}\left({\omega^{2}_{\rm p}\chi^{\rm
e}_{\alpha}{\mathcal E}_{x}+\omega_{\rm p}\chi^{\rm
e}_{\beta}\frac{\partial}{\partial t}{\mathcal E}_{x}+\chi^{\rm
e}_{\gamma}\frac{\partial^{2}}{\partial
t^{2}}{\mathcal E}_{x}}\right),                           \nonumber \\
\frac{\partial^{2}}{\partial t^{2}}{\mathcal
M}_{y}+\Gamma\frac{\partial}{\partial t}{\mathcal
M}_{y}+\omega^{2}_{0}{\mathcal M}_{y}=\omega^{2}_{\rm p}\chi^{\rm
m}_{\alpha}{\mathcal H}_{y}+\omega_{\rm p}\chi^{\rm
m}_{\beta}\frac{\partial}{\partial t}{\mathcal H}_{y}+\chi^{\rm
m}_{\gamma}\frac{\partial^{2}}{\partial t^{2}}{\mathcal H}_{y},
\label{eq2}
\end{eqnarray}
where $\chi _{\alpha }^{\rm e,m}$, $\chi _{\beta }^{\rm e,m}$ and
$\chi _{\gamma }^{\rm e,m}$ represent, respectively, the coupling
of the electric (magnetic) field and its first and second time
derivatives to the local electric (magnetic) dipole moments.
$\omega _{\rm p}$, $\Gamma ^{\rm e}$, $\Gamma ^{\rm m}$ and
$\omega _{0}$ can be viewed as the plasma frequency, damping
frequency and resonance frequency of the electric (magnetic)
dipole oscillators, respectively. Thus the frequency-domain
electric and magnetic susceptibilities are given
\begin{equation}
\chi ^{\rm e}\left( \omega \right)=\frac{\left( \omega _{\rm
p}^{\rm e}\right) ^{2}\chi _{\alpha }^{\rm e}+i\omega \omega _{\rm
p}^{\rm e}\chi _{\beta }^{\rm e}-\omega ^{2}\chi _{\gamma }^{\rm
e}}{-\omega ^{2}+i\omega \Gamma ^{\rm e}+\left( \omega _{0}^{\rm
e}\right) ^{2}}, \quad   \chi ^{\rm m}\left( \omega
\right)=\frac{\left( \omega _{\rm p}^{\rm m}\right) ^{2}\chi
_{\alpha }^{\rm m}+i\omega \omega _{\rm p}^{\rm m}\chi _{\beta
}^{\rm m}-\omega ^{2}\chi _{\gamma }^{\rm m}}{-\omega ^{2}+i\omega
\Gamma ^{\rm m}+\left( \omega _{0}^{\rm m}\right) ^{2}},
\label{eq41}
\end{equation}
So, the refractive index squared in the 2TDLM model reads
\begin{equation}
n^{2}\left( \omega \right) =1+\chi ^{\rm e}\left( \omega \right)
+\chi ^{\rm m}\left( \omega \right) +\chi ^{\rm e}\left( \omega
\right) \chi ^{\rm m}\left( \omega \right).
\label{eq3}
\end{equation}
Thus it follows from (\ref{eq1}) that the {\it photon effective
rest mass squared} in 2TDLM model is of the form
\begin{equation}
m_{\rm eff}^{2}=\frac{\hbar ^{2}\omega^{2}}{c^{4}}\left[{\chi
^{\rm e}\left( \omega \right) +\chi ^{\rm m}\left( \omega \right)
+\chi ^{\rm e}\left( \omega \right) \chi ^{\rm m}\left( \omega
\right)}\right]. \label{eq42}
\end{equation}
As an illustrative example, we apply Eq.(\ref{eq42}) to the case
of left-handed media, which is a kind of artificial composite
metamaterials with negative refractive index in the microwave
frequency region ($10^{9}$ Hz). More recently, this medium
captured considerable attention in various fields such as
condensed matter physics, materials science, applied
electromagnetics and
optics\cite{Smith,Veselago,Klimov,Pendry3,Shelby,Ziolkowski2}. The
most characteristic features of left-handed media are: ({\rm i})
both the electric permittivity $\epsilon$ and the magnetic
permeability $\mu$ are negative; ({\rm ii}) the Poynting vector
and wave vector of electromagnetic wave propagating inside it
would be antiparallel, {\it i.e.}, the wave vector {\bf {k}}, the
electric field {\bf {E}} and the magnetic field {\bf {H}} form a
{\it left-handed} system; ({\rm iii}) a number of peculiar
electromagnetic and optical properties such as the reversal of
both the Doppler shift and the Cherenkov radiation, anomalous
refraction, modified spontaneous emission rates and even reversals
of radiation pressure to radiation tension\cite{Veselago,Klimov}
arise. All these dramatically different propagation
characteristics stem from the sign change of the group velocity.
In what follows we calculate the {\it photon effective rest mass
squared} by making use of Eq.(\ref{eq42}) and the expressions
\begin{equation}
\epsilon \left( \omega \right) =1-\frac{\omega _{\rm
p}^{2}}{\omega \left( \omega +i\gamma \right) },    \quad  \mu
\left( \omega \right) =1-\frac{F\omega ^{2}}{\omega ^{2}-\omega
_{0}^{2}+i\omega \Gamma }
\end{equation}
for the effective dielectric parameter and magnetic permeability,
where the plasma frequency $\omega _{\rm p}$ and the magnetic
resonance frequency $\omega _{0}$ are in the GHz range ({\it
e.g.}, $\omega _{\rm p}=10.0$ GHz, $\omega _{0}=4.0$ GHz
\cite{Ruppin}), the damping parameters $\gamma =0.03\omega _{\rm
p}$, $\Gamma =0.03\omega _{0}$. The parameter $F$ may often be
chosen $0<F<1$, for example, $F=0.56$\cite{Ruppin}. By using
$n^{2}\left( \omega \right) =\epsilon \left( \omega \right) \mu
\left( \omega \right) $, we obtain\cite{Scr}
\begin{equation}
n^{2}\left( \omega \right) =1-\frac{\omega _{\rm p}^{2}}{\omega
^{2}}\left( 1+\frac{i\gamma }{\omega }\right) ^{-1}+\frac{F\omega
_{\rm p}^{2}}{\omega ^{2}}\left( 1+\frac{i\gamma }{\omega }\right)
^{-1}\left( 1-\frac{\omega _{0}^{2}}{\omega ^{2}}+\frac{i\Gamma
}{\omega }\right) ^{-1}-F\left( 1-\frac{\omega _{0}^{2}}{\omega
^{2}}+\frac{i\Gamma }{\omega }\right) ^{-1}. \label{eq57}
\end{equation}
It follows that the mass squared $m_{\rm eff}^{2}$ is obtained
\begin{equation}
m_{\rm eff}^{2}=\frac{\hbar ^{2}}{c^{4}}\left[{\omega _{\rm
p}^{2}\left( 1+\frac{i\gamma }{\omega }\right) ^{-1}-F\omega _{\rm
p}^{2}\left( 1+\frac{i\gamma }{\omega }\right) ^{-1}\left(
1-\frac{\omega _{0}^{2}}{\omega ^{2}}+\frac{i\Gamma }{\omega
}\right) ^{-1}+F\omega ^{2}\left( 1-\frac{\omega _{0}^{2}}{\omega
^{2}}+\frac{i\Gamma }{\omega }\right) ^{-1}}\right ].
\end{equation}
Note that for the electron-plasma medium, where the
electromagnetic parameters $\gamma=\Gamma=0$, $F=0$, one can
arrive at the following familiar formula
\begin{equation}
m_{\rm eff}^{2}=\frac{\hbar ^{2}\omega _{\rm p}^{2}}{c^{4}}
\label{eq19}
\end{equation}
with the plasma frequency squared $\omega _{\rm
p}^{2}=\frac{Ne^{2}}{\epsilon _{0}m_{\rm e}}$ (here $e$, $m_{\rm
e}$ and $N$ respectively denote the electron charge, mass and
electron number density in this electron plasma).

It should be emphasized that there is a difference in the
definition of {\it photon effective rest mass} between two types
of experimental schemes of testing photon mass\footnote{These two
types of experimental schemes are as follows: one is based on the
Hamiltonian density of electrodynamics and the other based on
Maxwellian equations.}. In the discussion of wave dispersion and
potential variations of the speed of light with frequency, which
are based on the Maxwellian equations ({\it rather than the
Hamiltonian or Lagrangian density of interacting electromagnetic
system}), the effective rest mass squared of photons is expressed
by Eq.(\ref{eq19}). Historically, this problem has been discussed
by many investigators, {\it e.g.}, Feinberg who considered the
possibility of the variation of the speed of light with frequency
in the sharply defined optical and radio pulses from pulsars and
took into account in more detail the observed variation of arrival
time with frequency for the radio waves attributed to the
interaction with interstellar electrons\cite{Feinberg}. However,
for those experimental schemes of testing photon rest mass, which
are based on the Hamiltonian density ({\it rather than the
Maxwellian equations}), the effective rest mass of photons due to
media dispersion ({\it i.e.}, the interaction of wave with charged
particles in environmental dilute plasma such as the secondary
cosmic rays) is two times that expressed in Eq.(\ref{eq19}). The
torsion balance experiment is just this type of schemes, which
will be considered in what follows.
\section{Effects of self-induced charge currents in the torsion balance experiment}
In this section, we consider the potential effects of self-induced
charge currents arising in the torsion balance
experiments\cite{Luo,Lakes}. The Lagrangian density of
electrodynamics reads
\begin{equation}
{\mathcal
L}=-\frac{1}{4}F_{\mu\nu}F_{\mu\nu}-\frac{1}{2}\mu^{2}_{\gamma}A_{\mu}A_{\mu}+\mu_{0}J_{\mu}A_{\mu},
\label{eq21}
\end{equation}
where
$\mu_{\gamma}^{2}=\left({\frac{m_{\gamma}c}{\hbar}}\right)^{2}$
with $m_{\gamma}$, $\hbar$ and $c$ being the photon rest mass,
Planck's constant and speed of light in a free vacuum,
respectively, and $\mu_{0}$ denotes the magnetic permeability in a
vacuum. It follows from (\ref{eq21}) that the canonical momentum
density is written
\begin{equation}
\pi_{\mu}=\frac{\partial{\mathcal L}}{\partial\dot{A}_{\mu}},
\quad               \vec{\pi}(x)=-{\bf E}(x).
\end{equation}
In the electron plasma, due to the conservation law of the
canonical momentum density, {\it i.e.}, $\frac{{\rm d}}{{\rm
d}t}(m_{\rm e}{\bf v}+e{\bf A})=0$, we have ${\bf
v}=-\frac{e}{m_{\rm e}}{\bf A}+{\bf C}$ with ${\bf C}$ being a
constant velocity. By using the formula ${\mathcal H}=-{\bf
E}\cdot{\dot{\bf A}}-{\mathcal L}$ and electric current density
${\bf J}=Ne{\bf v}=-\frac{N{e}^{2}}{m_{\rm e}}{\bf A}+Ne{\bf C}$,
one can arrive at (in SI)
\begin{equation}
{\mathcal H}=\frac{1}{2\mu_{0}}\left[{\frac{{\bf
E}^{2}}{c^{2}}+{\bf B}^{2}+\mu_{\gamma}^{2}{\bf
A}^{2}+2\frac{\mu_{0}N{e}^{2}}{m_{\rm e}}{\bf
A}^{2}+\frac{1}{\mu_{\gamma}^{2}c^{2}}\left(\nabla\cdot{\bf
E}-\frac{{\bf J}_{0}}{\epsilon_{0}}\right)^{2}}\right]+{\bf
J}_{0}{\bf A}_{0}-Ne{\bf C}\cdot{\bf A},
\label{eq23}
\end{equation}
where ${\bf J}_{0}=\rho$ and ${\bf A}_{0}=\phi$ are respectively
the electric charge density and electric scalar potential, and
$\epsilon_{0}$ represents the electric permittivity in a vacuum
and $c=\frac{1}{\sqrt{\epsilon_{0}\mu_{0}}}$. Note that here
$\frac{1}{\mu_{\gamma}^{2}c^{2}}\left(\nabla\cdot{\bf
E}-\frac{{\bf J}_{0}}{\epsilon_{0}}\right)^{2}$ results from both
the mass term ${\frac{\mu_{\gamma}^{2}}{c^{2}}}{\bf A}_{0}^{2}$
and the Gauss's law $\nabla\cdot{\bf E}=-\mu_{\gamma}^{2}{\bf
A}_{0}+\frac{{\bf J}_{0}}{\epsilon_{0}}$.

It follows from (\ref{eq23}) that the {\it total effective rest
mass squared} of electromagnetic fields in electron plasma is
given as follows
\begin{equation}
\mu_{\rm tot}^{2}=\mu_{\gamma}^{2}+2\frac{\mu_{0}N{e}^{2}}{m_{\rm
e}}.                                          \label{eq24}
\end{equation}
It is worthwhile to point out that according to the
Amp\`{e}re-Maxwell-Proca equation $\nabla\times{\bf
B}=\mu_{0}\epsilon_{0}\frac{\partial}{\partial t }{\bf
E}+\mu_{0}{\bf J}-\mu^{2}_{\gamma}{\bf A}$ and the consequent
$\nabla\times{\bf B}=\mu_{0}\epsilon_{0}\frac{\partial}{\partial t
}{\bf E}-[\mu^{2}_{\gamma}+\frac{\mu_{0}N{e}^{2}}{m_{\rm e}}]{\bf
A}+\mu_{0}Ne{\bf C}$, the $\mu_{\rm tot}^{2}$ should be $\mu_{\rm
tot}^{2}=\mu_{\gamma}^{2}+\frac{\mu_{0}N{e}^{2}}{m_{\rm e}}$
rather than that in (\ref{eq24}). This minor difference between
these two $\mu_{\rm tot}^{2}$ results from the derivative
procedure applied to the Lagrangian density by using the
Euler-Lagrange equation. Although it seems from the
Amp\`{e}re-Maxwell-Proca equation that $\mu_{\rm
tot}^{2}=\mu_{\gamma}^{2}+\frac{\mu_{0}N{e}^{2}}{m_{\rm e}}$, the
factor-$2$ in (\ref{eq24}) cannot be ignored in calculating the
torque acting on the torsion balance due to the photon effective
rest mass, which will be confirmed in what follows. In the
classical electromagnetics, the following familiar formulae are
given
\begin{equation}
\nabla\times{\bf B}=\mu_{0}{\bf J},  \quad   m_{d}=\pi r^{2}I,
\quad          I={\bf J}\cdot {\bf S},  \quad
\vec{\tau}={\bf m}_{d}\times {\bf B} \label{eq25}
\end{equation}
and
\begin{equation}
\nabla\times{\bf A}={\bf B},  \quad  a_{d}=\pi
r^{2}\frac{\Phi}{\mu_{0}},  \quad   \Phi={\bf B}\cdot{\bf S},
\label{eq26}
\end{equation}
where $\nabla\times{\bf B}=\mu_{0}{\bf J}$ holds when the electric
field strength ${\bf E}$, electric current density ${\bf J}$ are
vanishing in the Amp\`{e}re-Maxwell equation. In Eq.(\ref{eq25}),
$m_{d}=\pi r^{2}I$ means that a current loop of radius $r$
carrying current $I$ gives rise to a magnetic dipole moment
$m_{d}$; ${\bf S}$ in $I={\bf J}\cdot {\bf S}$ denotes the area
vector through which the electric current density vector ${\bf J}$
penetrates; $\vec{\tau}={\bf m}_{d}\times {\bf B}$ means that an
electric-current loop with a magnetic dipole moment ${\bf m}_{d}$,
immersed in a magnetic field ${\bf B}$, experiences a torque
$\vec{\tau}={\bf m}_{d}\times {\bf B}$. Eqs.(\ref{eq26}) means in
the torsion balance experiment, a toroid coil contains a loop of
magnetic flux ${\Phi}$ which acts as a dipole source $a_{d}$ of
magnetic vector potential via $\nabla\times{\bf A}={\bf B}$ with
the magnetic field ${\bf B}$ within the toroid as the source term.

It follows from (\ref{eq23}) that in the Hamiltonian density
${\mathcal H}$ of electromagnetic system, there are
$\frac{1}{2}{\bf B}^{2}$ and
$\frac{1}{2}\left(\mu_{\gamma}^{2}+2\frac{\mu_{0}N{e}^{2}}{m_{\rm
e}}\right){\bf A}^{2}$ ({\it i.e.}, $\frac{1}{2}\left(\mu_{\rm
tot}{\bf A}\right)^{2}$), which provide us with some useful
insights into calculating the torque produced by the interaction
of magnetic dipole vector potential moment ${\bf a}_{d}$ with the
ambient cosmic magnetic vector potential acting upon the toroid in
the torsion balance experiment. For convenience, the Equations
(\ref{eq26}) may be rewritten as follows
\begin{equation}
\nabla\times(\mu_{\rm tot}{\bf A})=\mu_{0}\left(\frac{\mu_{\rm
tot}{\bf B}}{\mu_{0}}\right),     \quad        \mu_{\rm
tot}a_{d}=\pi r^{2}\left(\frac{\mu_{\rm tot}
\Phi}{\mu_{0}}\right),                       \quad \frac{\mu_{\rm
tot}\Phi}{\mu_{0}}=\left(\frac{\mu_{\rm tot}{\bf
B}}{\mu_{0}}\right)\cdot{\bf S}.                    \label{eq27}
\end{equation}

In order to calculate the torque acting on the torsion balance, we
compare the equations and expressions in (\ref{eq27}) with those
in (\ref{eq25}) as follows:
$$
\vbox{\tabskip=0pt \offinterlineskip \halign to \hsize {\strut#&
\vrule#\tabskip=1em plus2em &\hfil#\hfil& \vrule# &\hfil#\hfil&
\vrule#&\hfil#\hfil & \vrule#\tabskip=0pt\cr \noalign{\hrule}

&& \omit\hidewidth  \hidewidth & & \omit\hidewidth the familiar
case \hidewidth && \omit\hidewidth the case due to photon mass
\hidewidth & \cr \noalign{\hrule} && in ${\mathcal H}$ &&
$\frac{1}{2}{\bf B}^{2}$ &&$\frac{1}{2}\left(\mu_{\rm tot}{\bf
A}\right)^{2}$&  \cr \noalign{\hrule} && Equations &&
$\nabla\times{\bf B}=\mu_{0}{\bf J}$ && $\nabla\times(\mu_{\rm
tot}{\bf A})=\mu_{0}\left(\frac{\mu_{\rm tot}{\bf
B}}{\mu_{0}}\right)$& \cr \noalign{\hrule} && dipole moment &&
$m_{d}=\pi r^{2}I$ && $ \mu_{\rm tot}a_{d}=\pi
r^{2}\left(\frac{\mu_{\rm tot} \Phi}{\mu_{0}}\right)$& \cr
\noalign{\hrule} && flux && $I={\bf J}\cdot {\bf S}$ && $\quad
\frac{\mu_{\rm tot}\Phi}{\mu_{0}}=\left(\frac{\mu_{\rm tot}{\bf
B}}{\mu_{0}}\right)\cdot{\bf S}$& \cr \noalign{\hrule} && torque
&& $\vec{\tau}={\bf m}_{d}\times {\bf B}$ && $\vec{\tau}=?$ &  \cr
\noalign{\hrule} \noalign{\smallskip}
 \multispan7*{Table {\it 1}}
\hfil\cr}}
$$

Thus it follows from the analogies in the above table that,
$?=(\mu_{\rm tot}{\bf a}_{d})\times(\mu_{\rm tot}{\bf A}) $, {\it
i.e.}, the torque $\vec{\tau}=\mu_{\rm tot}^{2}{\bf
a}_{d}\times{\bf A}$. Here the torque $\vec{\tau}$ arising from
the interaction between the dipole field of magnetic potentials
(resulting from the electric current $I$ in the toroid windings)
with the ambient vector potential ${\bf A}$ acts on the torsion
balance. It should be noted again that here $\mu_{\rm
tot}^{2}=\mu_{\gamma}^{2}+2\frac{\mu_{0}N{e}^{2}}{m_{\rm e}}$
rather than $\mu_{\rm
tot}^{2}=\mu_{\gamma}^{2}+\frac{\mu_{0}N{e}^{2}}{m_{\rm e}}$.
\\

Note that there exists a term $\mu_{0}Ne{\bf C}$ in the
Amp\`{e}re-Maxwell-Proca equation $\nabla\times{\bf
B}=\mu_{0}\epsilon_{0}\frac{\partial}{\partial t }{\bf
E}-[\mu^{2}_{\gamma}+\frac{\mu_{0}N{e}^{2}}{m_{\rm e}}]{\bf
A}+\mu_{0}Ne{\bf C}$. Does this term has influences on the torsion
balance? In what follows we will discuss this problem.

In the similar fashion, we consider the familiar Hamiltonian
density $-{\bf M}\cdot{\bf B}$, which describes the interaction of
the magnetic moments with the magnetic field ${\bf B}$. According
to the Amp\`{e}re-Maxwell-Proca equation with
$\mu_{0}\epsilon_{0}\frac{\partial}{\partial t }{\bf E}$ and
$\frac{\mu_{0}N{e}^{2}}{m_{\rm e}}{\bf A}$ being ignored, we have
$\nabla\times{\bf B}=\mu_{0}Ne{\bf C}$, which can be rewritten as
$\nabla\times{\bf B}=\left(\mu_{0}\mu_{\rm
tot}\right)\left(\frac{Ne{\bf C}}{\mu_{\rm tot}}\right)$. In the
meanwhile,  ${\mathcal H}=-Ne{\bf C}\cdot{\bf A}$ can also be
rewritten as ${\mathcal H}=-\left(\frac{Ne{\bf C}}{\mu_{\rm
tot}}\right)\cdot\left(\mu_{\rm tot}{\bf A}\right)$.

It is known that in the sufficiently large space filled with
homogeneous magnetic moments with volume density (magnetization)
being ${\bf M}$, the interior magnetic field strength ${\bf B}$
produced by homogeneously distributed magnetic moments is just
$\mu_{0}{\bf M}$\footnote{This may be considered in two ways:
({\rm i}) if the direction of ${\bf M}$ is assumed to be parallel
to the third component of Cartesian coordinate system, then the
potentially non-vanishing components of ${\bf B}$ and ${\bf H}$
are also the third ones. It follows from the Amp\`{e}re's law
$\nabla\times{\bf H}={\bf J}_{\rm free}$ with ${\bf J}_{\rm
free}=0$ that $H_{3}$ is constant, and $H_{3}$ can be viewed as
zero. So, according to the expression ${\bf B}=\mu_{0}({\bf
H+M})$, we obtain ${\bf B}=\mu_{0}{\bf M}$; ({\rm ii}) In
accordance with Amp\`{e}re's circuital law, the magnetic induction
$B$ inside a long solenoid with $N$ circular loops carrying a
current $I$ in the region remote from the ends is axial, uniform
and equal to $\mu_{0}$ times the number of Amp\`{e}re-turns per
meter $nI$, {\it i.e.}, $B=\mu_{0}nI$, where $n=\frac{N}{l}$ with
$l$ being the solenoid length. Let $S$ denote the cross-section
area of this long solenoid. The volume density of magnetic moments
$IS$ resulting from the carried current $I$ is
$M=\frac{(IS)nl}{V}$ with the solenoid volume $V=Sl$. This,
therefore, means that the volume density of magnetic moments just
equals the number of Amp\`{e}re-turns per meter $nI$, {\it i.e.},
$M=nI$. So, we have $B=\mu_{0}M$.}, {\it i.e.}, $\nabla\times{\bf
A}=\mu_{0}{\bf M}$ or $\nabla\times\left({\mu_{\rm tot}}{\bf
A}\right)=\left(\mu_{0}{\mu_{\rm tot}}\right){\bf M}$.

Thus the comparisons between ``{\it the familiar case}'' and {\it
the case due to the constant} ${\bf C }$ are illustrated in the
following table,
$$
\vbox{\tabskip=0pt \offinterlineskip \halign to \hsize {\strut#&
\vrule#\tabskip=1em plus2em &\hfil #\hfil & \vrule#&\hfil#\hfil&
\vrule#&\hfil#\hfil & \vrule#\tabskip=0pt\cr \noalign{\hrule}

&& \omit\hidewidth  \hidewidth & & \omit\hidewidth the familiar
case \hidewidth && \omit\hidewidth the case due to constant ${\bf
C }$ \hidewidth & \cr \noalign{\hrule} && in ${\mathcal H}$ &&
$\frac{1}{2}{\bf B}^{2}$ &&$\frac{1}{2}\left(\mu_{\rm tot}{\bf
A}\right)^{2}$&  \cr \noalign{\hrule} && interaction Hamiltonian
density && ${\mathcal H}=-{\bf M}\cdot{\bf B}$ && ${\mathcal
H}=-\left(\frac{Ne{\bf C}}{\mu_{\rm
tot}}\right)\cdot\left(\mu_{\rm tot}{\bf A}\right)$& \cr
\noalign{\hrule} && Equations &&$\nabla\times\left({\mu_{\rm
tot}}{\bf A}\right)=\left(\mu_{0}{\mu_{\rm tot}}\right){\bf M}$ &&
$\nabla\times{\bf B}=\left(\mu_{0}\mu_{\rm
tot}\right)\left(\frac{Ne{\bf C}}{\mu_{\rm tot}}\right)$&  \cr
\noalign{\hrule} && torque && $\vec{\tau}={\bf M}\times{\bf B}$&&
$\vec{\tau}=?$& \cr \noalign{\hrule}  \noalign{\smallskip}
 \multispan7*{Table {\it 2}}
\hfil\cr}}
$$

It follows that $?=\left(\frac{Ne{\bf C}}{\mu_{\rm
tot}}\right)\times\left(\mu_{\rm tot}{\bf A}\right)$, {\it i.e.},
$\vec{\tau}=Ne{\bf C}\times{\bf A}$, which dose not act on the
toroid in the torsion balance experiment.
\\

Since we have discussed the related preliminary preparation for
the effects of self-induced charge currents, we think the order of
magnitude of its effects in experiments ({\it e.g.}, the extra
torque due to the photon effective mass acting on the toroid in
the torsion balance experiment) deserves evaluation.
\section{Discussion and remarks: self-induced charge currents in dilute plasma}
Here we discuss the effective rest mass of photon resulting from
the self-induced charge currents in the torsion balance
experiments\cite{Luo,Lakes}. The self-induced charge current
arises mainly from the two sources around us: ({\rm i}) muon
($\mu$) component and alpha-particles in cosmic rays; ({\rm ii})
decay daughter (alpha-particles) of radioactive radon gas
($^{222}_{86}{\rm Rn}$, $^{220}_{86}{\rm Rn}$) in room
environment.

It is known that at the sea level, the current density of muon
component in secondary cosmic rays is about $1\times 10^{-2}{\rm
cm}^{-2}\cdot{\rm s}^{-1}$. Assuming that the muon velocity
approaches speed of light, the volume density of muon can be
derived and the result is $N_{\mu}=0.3\times 10^{-6}$ ${\rm
m}^{-3}$. So, according to Eq.(\ref{eq24}) ({\it i.e.}, $m_{\rm
eff}=\frac{\hbar}{c}\sqrt{\frac{2Ne^{2}}{\epsilon_{0}m_{\mu}c^{2}}}$),
electromagnetic wave with wavelength\footnote{Since the mean
distance of two muons in secondary cosmic rays at sea level is
about 100 m, Lorentz's mean-field formulation (Lorentz's
dispersion theory, Lorentz's electron theory, 1895) is not
applicable to the electromagnetic wave with wavelength $\lambda
\ll 100$ m.  So, the above effective mass formula is no longer
valid for this case ({\it i.e.}, $\lambda \ll 100$ m). In fact,
the electromagnetic wave with wavelength $\lambda \ll 100$ m dose
not acquire this effective rest mass. But the ambient cosmic
magnetic vector potentials (interstellar magnetic fields) with low
(or zero) frequencies will truly acquire this effective rest
mass.} $\lambda \gg 100$ m at the sea level acquires an effective
rest mass about $0.6\times 10^{-53}$ Kg. Note that the
above-mentioned muon volume density ($0.3\times 10^{-6}{\rm
m}^{-3}$) is the datum at the sea level. In the Luo's rotating
torsion balance experiment whose total apparatus is located in
their cave laboratory, on which the least thickness of the cover
is more than $40$ m\cite{Luo}, it is reasonably believed that the
muon density of cosmic rays in the vacuum chamber of Luo's
experiment may be one or two orders of magnitude\footnote{Readers
may be referred to the following data of muon current density in
underground cosmic rays: the muon current densities are
$10^{-4}{\rm cm}^{-2}\cdot{\rm s}^{-1}$ and $10^{-6}{\rm
cm}^{-2}\cdot{\rm s}^{-1}$ at the equivalent water depth of $100$
m and $1000$ m under the ground, respectively.} less than that at
the sea level, and therefore the effective rest mass acquired by
photons in the vacuum chamber of Luo's torsion balance experiment
may be about $(0.6\sim 2)\times 10^{-54}$ Kg.

It is believed that the air molecules in the vacuum chamber of
torsion balance experiment cannot be ionized by the
alpha-particles of cosmic rays, because the mean free path of
alpha-particle moving at about $10^{7}{\rm m}\cdot{\rm s}^{-1}$ in
the dilute air with the pressure being only $10^{-2}$ Pa\cite{Luo}
is too large (more than $10^{6}$ m)\footnote{According to some
handbooks, the free paths of alpha-particles in air (with pressure
$10^{5}$ Pa) are 2.5, 3.5, 4.6, 5.9, 7.4, 8.9, 10.6 cm,
corresponding respectively to the energy 4.0, 5.0, 6.0, 7.0, 8.0,
9.0, 10.0 Mev. So, in the vacuum chamber with pressure $10^{-2}$
Pa in Luo's experiment, the free path of alpha-particles (with
energy much more than 10 Mev, even perhaps more than 100 Mev) in
secondary cosmic rays is more than $10^{7}$ times several cm,
which means that the air molecules in the low-pressure vacuum
chamber cannot be easily ionized by the alpha-particles of cosmic
rays.}. So, the air medium has nothing to contribute to the
effective rest mass of photons. As far as the alpha-particle
itself is concerned, it follows, from the fact that at the sea
level the alpha-particle number density in secondary cosmic rays
is about one order of magnitude less than that of muon, that the
alpha-particle in cosmic rays itself contributes about $10^{-54}$
Kg to photon rest mass.

Radon ($_{86}{\rm Rn}$) possesses the $\alpha$-radioactivity.
Generally speaking, the Rn density in room environment is
$10^{-8}{\rm m}^{-3}$ or so, which may be the same order of
magnitude as particle density in secondary cosmic
rays\footnote{The effective doses of cosmic rays and Rn humans
suffer are respectively $0.4$ mSv (typical range: 0.3-1.0 mSv) and
$1.2$ mSv (typical range: 0.2-10 mSv).}, and the decay daughter
(alpha-particles) will therefore give rise to a photon effective
mass of about $10^{-54}$ Kg.
\\

Thus the effective rest mass induced by muons and alpha-particles
in secondary cosmic rays and radon gas can be compared to the new
upper limit ($1.2\times 10^{-54}$ Kg) obtained by Luo {\it et al.}
in their recent rotating torsion balance experiment\cite{Luo}. So,
as far as the contribution of ions of cosmic rays to the photon
rest mass is concerned, this upper limit ($1.2\times 10^{-54}$ Kg)
becomes just a critical value. We think Luo's experimentally
obtained upper limit is of much interest and therefore deserves
further experimental investigation to get a more precise upper
limit on photon mass.
\section{Conclusion}
In this paper,
\\

{\rm i}) it is verified that the expression (\ref{eq1}), {\it
i.e.}, $\frac{m_{\rm eff}^{2}c^{4}}{\hbar ^{2}}=(1-n^{2})\omega
^{2}$ can be applied to calculating the effective rest mass of
photons in the artificial composite electromagnetic metamaterials
with complicated permittivity and permeability.
\\

{\rm ii}) it is shown that the extra torque
$2\frac{\mu_{0}Ne^{2}}{m}{\bf a}_{d}\times {\bf A}$ by the
self-induced charge currents will also acts upon the toroid in the
torsion balance experiment whose apparatus is immersed in the
dilute plasma such as the cosmic rays and radon gas. Here the
photon effective rest mass squared is $\mu_{\rm
eff}^{2}=2\frac{\mu_{0}N{e}^{2}}{m_{\rm e}}$ rather than $\mu_{\rm
eff}^{2}=\frac{\mu_{0}N{e}^{2}}{m_{\rm e}}$.

The following difference between these two $\mu_{\rm eff}^{2}$
should be pointed out: the theoretical mechanism of torsion
balance experiment lies in the Lagrangian or Hamiltonian
(\ref{eq23}) of interacting electromagnetic system, so, the
effective rest mass squared of electromagnetic fields is $\mu_{\rm
eff}^{2}=2\frac{\mu_{0}N{e}^{2}}{m_{\rm e}}$. However, for those
experimental schemes\cite{Feinberg,Fischbach,Chernikov} of photon
rest mass stemming from the Amp\`{e}re-Maxwell-Proca equation
$\nabla\times{\bf B}=\mu_{0}\epsilon_{0}\frac{\partial}{\partial t
}{\bf E}-[\mu^{2}_{\gamma}+\frac{\mu_{0}N{e}^{2}}{m_{\rm e}}]{\bf
A}+\mu_{0}Ne{\bf C}$, the effective rest mass squared of
electromagnetic fields is $\mu_{\rm
eff}^{2}=\frac{\mu_{0}N{e}^{2}}{m_{\rm e}}$ rather than $\mu_{\rm
eff}^{2}=2\frac{\mu_{0}N{e}^{2}}{m_{\rm e}}$. The so-called
experimental realizations stemming from the
Amp\`{e}re-Maxwell-Proca equation are as follows: pulsar test of a
variation of the speed of light with frequency\cite{Feinberg};
geomagnetic limit on photon mass based on the analysis of
satellite measurements of the Earth's field\cite{Fischbach};
experimental test of Amp\`{e}re's law at low
temperature\cite{Chernikov}, {\it etc.}. In all these experimental
schemes, the effective mass squared is $\mu_{\rm
eff}^{2}=\frac{\mu_{0}N{e}^{2}}{m_{\rm e}}$.
\\

{\rm iii}) the self-induced charge currents in dilute plasma
presented above ({\it i.e.}, muons and alpha-particles in
secondary cosmic rays and radon gas) will give arise to an
effective rest mass of about $10^{-54}$ Kg, which can be compared
to the new upper limit on photon rest mass obtained in Luo's
rotating torsion balance experiment.
\\

\textbf{Acknowledgements}  This project is supported in part by
the National Natural Science Foundation of China under the project
No. $90101024$.

\end{document}